\documentclass[sigconf]{acmart}

\acmSubmissionID{593}   

\setcitestyle{square}
\usepackage{stfloats}
\copyrightyear{2024}
\acmYear{2024}
\setcopyright{acmlicensed}\acmConference[SIGGRAPH Conference Papers '24]{Special Interest Group on Computer Graphics and Interactive Techniques Conference Conference Papers '24}{July 27-August 1, 2024}{Denver, CO, USA}
\acmBooktitle{Special Interest Group on Computer Graphics and Interactive Techniques Conference Conference Papers '24 (SIGGRAPH Conference Papers '24), July 27-August 1, 2024, Denver, CO, USA}
\acmDOI{10.1145/3641519.3657442}
\acmISBN{979-8-4007-0525-0/24/07}

\usepackage{graphicx}
\usepackage{soul}
\usepackage{amsmath}

\usepackage{amssymb}
\usepackage{microtype}
\usepackage{wrapfig}
\usepackage{xcolor}
\usepackage{booktabs}
\usepackage{multirow}
\usepackage{xspace}

\usepackage[group-separator={,}]{siunitx}

\newcommand{\eg}{e.g.}
\newcommand{\ie}{i.e.}
\newcommand{\etal}{et~al.\ }

\newcommand{\refSec}[1]{Sec.~\ref{sec:#1}}
\newcommand{\refFig}[1]{Fig.~\ref{fig:#1}}
\newcommand{\refEq}[1]{Eq.~\ref{eq:#1}}
\newcommand{\refTbl}[1]{Tbl.~\ref{tab:#1}}
\newcommand{\refAlg}[1]{Alg.~\ref{alg:#1}}

\newcommand{\refFigSupp}[1]{Suppl.~Fig.~\ref{supplemental-fig:#1}}

\newcommand{\change}[1]{#1}

\soulregister\ref7
\soulregister\cite7
\soulregister\refFig7
\soulregister\refAlg7
\soulregister\refTab7
\soulregister\refTbl7
\soulregister\refEq7
\soulregister\cite7
\soulregister\ref7
\soulregister\pageref7
\soulregister\shortcite7
\soulregister\textsubscript7
\soulregister\eg0
\soulregister\eg7
\soulregister\ie0
\soulregister\etal0
\soulregister\ac7
\soulregister\acp7
\soulregister\refFigSupp7
\soulregister\url7

\DeclareGraphicsExtensions{.png,.jpg,.pdf,.ai,.psd}
\usepackage{xr}
\makeatletter
\newcommand*{\addFileDependency}[1]{
  \typeout{(#1)}
  \@addtofilelist{#1}
  \IfFileExists{#1}{}{\typeout{No file #1.}}
}
\makeatother

\usepackage[nolist]{acronym}
\begin{acronym}
\acro{AABB}{axis-aligned bounding box}
\acro{BVH}{bounding volume hierarchy}
\acro{CNN}{convolutional neural network}
\acro{DOF}{degrees-of-freedom}
\acro{FP}{false-positive}
\acro{FN}{false-negative}
\acro{KDOP}[$k$-DOP]{$k$ discretely-oriented polytope}
\acro{MC}{Monte Carlo}
\acro{MLP}{multi-layer perceptron}
\acro{NIF}{neural intersection function}
\acro{NN}{neural network}
\acro{OBB}{oriented bounding box}
\acro{SGD}{stochastic gradient descent}
\acro{TP}{true-positive}
\acro{TN}{true-negatives}
\acro{VAE}{variational auto-encoder}

\end{acronym}

\def\myparagraph#1{\paragraph{#1}}

\usepackage{flushend}
\usepackage{calc}
\usepackage{icomma}
\usepackage{algorithm}
\usepackage{algpseudocode}
\usepackage{kvsetkeys}
\usepackage{url}

\definecolor{AABoxColor}{HTML}{4285f4}
\definecolor{OBoxColor}{HTML}{ea4335}
\definecolor{SphereColor}{HTML}{fbbc04}
\definecolor{AAElliColor}{HTML}{34a853}
\definecolor{OElliColor}{HTML}{ff6d01}
\definecolor{kDOPColor}{HTML}{46bdc6}
\definecolor{OccNetColor}{HTML}{d611d6}
\definecolor{BVHColor}{HTML}{2030E0}
\definecolor{OurkDOPColor}{HTML}{B0B0B0}
\definecolor{OurReLUFieldColor}{HTML}{909090}
\definecolor{OurNNColor}{HTML}{606060}
\definecolor{OurNNEarlyColor}{HTML}{000000}

\usepackage{graphicx}
\usepackage{enumitem}
\usepackage{booktabs}
\usepackage{xspace}
\usepackage{comment}
\usepackage{algorithm}
\usepackage{algpseudocode}
\newcommand{\R}{\mathbb{R}}

\newcommand{\method}[1]{\texttt{\textcolor{#1Color}{\textbf{#1}}}\xspace}

\newcolumntype{P}{r<{\%}}

\newcommand{\parameters}{\ifmmode\theta\else$\theta$\xspace\fi}
\newcommand{\indicator}{\ifmmode f\else $f$\xspace\fi}
\newcommand{\location}{\ifmmode\mathbf x\else$\mathbf x$\xspace\fi}
\newcommand{\indicatorDimension}{\ifmmode n\else $n$\xspace\fi}
\newcommand{\query}{\ifmmode g\else$g$\xspace\fi}
\newcommand{\region}{\ifmmode\mathbf{r}\else$\mathbf{r}$\xspace\fi}
\newcommand{\queryDimension}{\ifmmode m\else$m$\xspace\fi}
\newcommand{\bounding}{\ifmmode {h_\theta}\else\ensuremath{{h_\theta}}\fi}
\newcommand{\cost}{\ifmmode c\else$c$\xspace$\xspace\fi}
\newcommand{\fnCost}{\ifmmode\alpha\else\ensuremath{\alpha}\fi}
\newcommand{\fpCost}{\ifmmode\beta\else\ensuremath{\beta}\fi}
\newcommand{\loss}{\ifmmode\mathcal{L}\else$\mathcal{L}$\xspace\fi}
\newcommand{\learningTime}{\ifmmode t\else$t$\xspace\fi}
\newcommand{\batchSize}{\ifmmode b\else$b$\xspace\fi}
\newcommand{\sampleCount}{\ifmmode n_\mathrm s\else$n_\mathrm s$\xspace\fi}
\newcommand{\samples}{\ifmmode \xi\else$\xi$\xspace\fi}
\newcommand{\learingRate}{\ifmmode\lambda\else$\lambda$\xspace\fi}
\newcommand{\supervision}{\ifmmode y\else$y$\xspace\fi}
\newcommand{\prediction}{\ifmmode\hat y\else$\hat y$\xspace\fi}
\newcommand{\nnloss}{\ifmmode\hat{\mathcal{L}}\else$\hat{\mathcal{L}}$\xspace\fi}
\newcommand{\combinedcost}{\ifmmode C\else$C$\xspace\fi}

\newcommand{\layer}{\ifmmode\mathsf A\else$\mathsf A$\xspace\fi}
\newcommand{\layerOne}{\ifmmode\layer_1\else$\layer_1$\xspace\fi}
\newcommand{\layerTwo}{\ifmmode\layer_2\else$\layer_2$\xspace\fi}
\newcommand{\layerThree}{\ifmmode\layer_3\else$\layer_3$\xspace\fi}
\newcommand{\nonlinearity}{\ifmmode\mathtt{nl}\else$\mathtt{nl}$\fi}

\newcommand{\numTests}{\ifmmode N\else\ensuremath{N}\fi}
\newcommand{\winner}[1]{\textbf{#1}}
\newcommand{\notAvailable}{ \multicolumn{1}{c}{---$^1$}}

\begin{document}

\title[Neural Bounding]{Neural Bounding}

\author{Stephanie Wenxin Liu}
\affiliation{%
	\institution{Birkbeck, University of London}
	\country{United Kingdom}
}
\email{wenxin.liu.cs@gmail.com}

\author{Michael Fischer}
\affiliation{%
	\institution{University College London}
	\country{United Kingdom}
}
\email{m.fischer@cs.ucl.ac.uk}

\author{Paul D. Yoo}
\affiliation{%
	\institution{Birkbeck, University of London}
	\country{United Kingdom}
}
\email{p.yoo@bbk.ac.uk}

\author{Tobias Ritschel}
\affiliation{%
	\institution{University College London}
	\country{United Kingdom}
}
\email{t.ritschel@ucl.ac.uk}

\begin{abstract}
Bounding volumes are an established concept in computer graphics and vision tasks but have seen little change since their early inception. 
In this work, we study the use of neural networks as bounding volumes.
Our key observation is that bounding, which so far has primarily been considered a problem of computational geometry, can be redefined as a problem of learning to classify space into free or occupied.
This learning-based approach is particularly advantageous in high-dimensional spaces, such as animated scenes with complex queries, where neural networks are known to excel.
However, unlocking neural bounding requires a twist: allowing -- but also limiting -- false positives, while ensuring that the number of false negatives is strictly zero.
We enable such tight and conservative results using a dynamically-weighted asymmetric loss function.
Our results show that our neural bounding produces up to an order of magnitude fewer false positives than traditional methods.
In addition, we propose an extension of our bounding method using early exits that accelerates query speeds by 25\,\%. We also demonstrate that our approach is applicable to non-deep learning models that train within seconds.
\change{Our project page is at \url{https://wenxin-liu.github.io/neural_bounding/}.}

\end{abstract}


\begin{teaserfigure}
    \includegraphics[width=\textwidth]{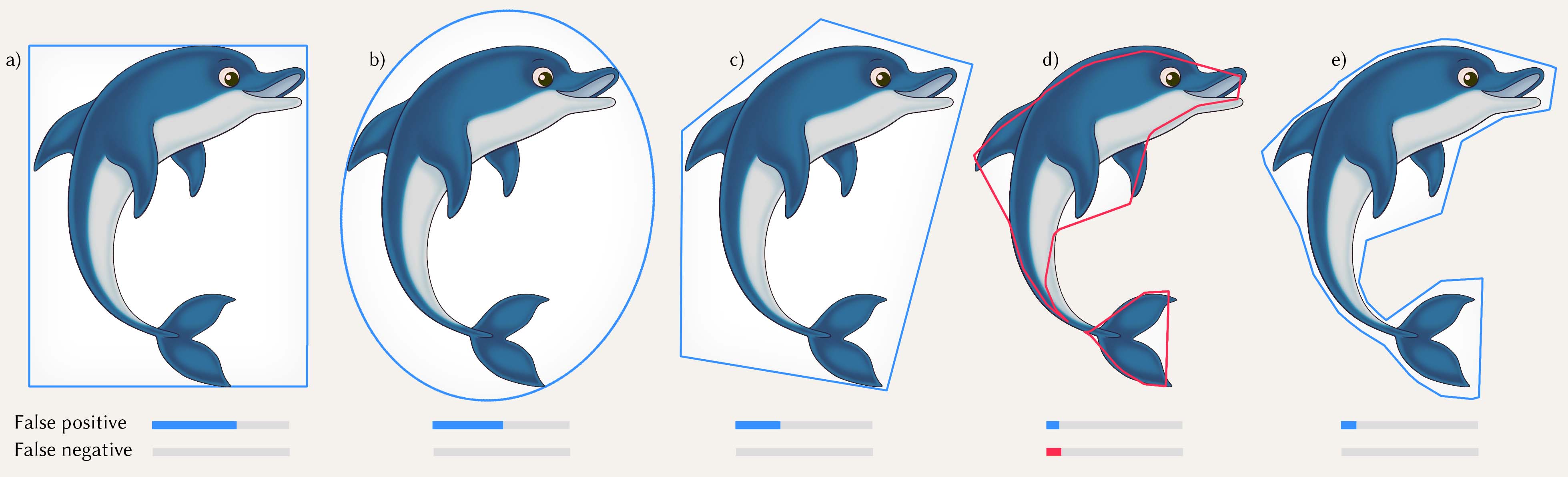}
    \vspace{-.4cm}
    \caption[]{
    Different bounding volume types classifying 2D space as maybe-object or certainly-not-object, from left to right: box (\textbf{a}), ellipsoid (\textbf{b}), $k$-oriented planes (\textbf{c}), common neural networks (\textbf{d}) and a neural network trained using our approach (\textbf{e}).
    While common boundings are not tight, common neural networks are not conservative, missing parts of the dolphin, while ours is both tight and has no false negatives.}
    \label{fig:Teaser}
\end{teaserfigure}
\maketitle

\section{Introduction}\label{sec:Introduction}
Efficiently testing two-, three- or higher-dimensional points or ranges for intersections with extended primitives is at the core of many interactive graphics tasks.
Examples include testing the 3D position of a particle in a fluid simulation against an animated character mesh, testing rays against a 3D medical scan volume or testing a drone's flight path against time-varying obstacles.

To accelerate all these queries, it is popular to use a hierarchy of tests: if intersection with a simple \emph{bounding primitive} -- such as a box -- that conservatively contains a more complex primitive fails, one can skip the costly test with the more complex primitive.
For a correct algorithm, the \ac{FN} rate of the first test must be zero, \ie, bounding must never miss a true positive intersection.

For efficiency, the main trade-offs are 
i) the cost of testing the bounding primitive, ii) the cost of intersecting the original primitive, and iii) the \ac{FP} rate of the bounding primitive. The \ac{FP} rate measures how often an initial positive intersection with the bounding primitive turns out to be negative upon more detailed testing with the original primitive, leading to wasted computation. A successful bounding method will have both a low testing cost and a low \ac{FP} rate.
Typical bounding solutions include spheres, boxes, oriented boxes or \acp{KDOP} \cite{ericson2004real}.
However, fitting those primitives, particularly in higher dimensions,  can result in poor \ac{FP} rates as they remain convex and further may require significant implementation effort \cite{schneider2002geometric}. 
In this article, we thus show how to train neural networks to unlock high-dimensional, non-linear, concave bounding with a combination of simplicity, flexibility and testing speed.

While the \ac{FP} rate is the main concern for \emph{efficiency}, for \emph{correctness} of the bounding algorithm, the challenge is to develop a \ac{NN} that is trained to produce bounds with strictly zero false negatives. This is crucial, as the \ac{FN} rate quantifies how often the algorithm erroneously classifies an actual intersection as non-intersection -- such misclassifications will result in truncated geometry features and cut-off object parts, as exemplified by the fins of the dolphin in \refFig{Teaser}, d. 
A straightforward solution would be to first find a bounding primitive and then compress it using a \ac{NN}.
Another approach would be to learn the \ac{NN} to approximate the complex primitive and later make the approximation conservative.
Instead, we show that with the right initialization and schedule for weighting \acp{FP} and \acp{FN}, \change{we can incentivize a neural network to become bounding.
While our method works well in many cases, we would like to point out that it is a heuristic and thus --not uncommon for neural networks-- does not provide strict guarantees.}

As it could appear that executing a neural network for testing bounds is too time-intensive to be useful, we carefully study architectures that are both small and simple (inspired by \cite{reiser2021kilonerf} or \cite{karnewar2022relu}), such that they are only slightly more expensive than linear ones or traditional intersection tests.
We further demonstrate that our approach is also amenable to optimizing non-neural representations, such as \acp{KDOP}.

We show application to two, three and 4D point queries, 2D and 3D range queries as well as queries of dynamic scenes, including scenes with multiple degrees of freedom, and compare these results to classic bounding methods, such as spheres, boxes and \acp{KDOP}.

\section{Previous work}\label{sec:PreviousWork}

Bounding $n$-D objects is a core operation in graphics.
Classic algorithms can be extremely straightforward, such as axis-aligned boxes, but already fitting a sphere can be more involved than it seems at first.
For an established textbook with many bounding and intersection algorithms, see \cite{schneider2002geometric}.

When it comes to bounding complex objects, the situation is more difficult:
For a single object that is dominantly convex, \acp{KDOP} have shown useful \cite{kay1986ray, klosowski1998efficient}.
Another option is to perform convex object-decomposition \cite{bergen1997efficient, ehmann2001accurate}.
3D ray-queries, are the most relevant for such tests, and typically performed on hierarchies, \eg, \cite{gu2013efficient, gunther2007realtime}; for a survey see Meister \etal \shortcite{meister2021survey}. 

Recently, \acp{NN} have changed many operations in graphics, but notably not bounding.
\Acp{NIF} \cite{fujieda2023neural} predict intersections and could also be used to intersect boundings, but do not attempt to be conservative, which allows the possibility to miss parts of the object they enclose.
Moreover, \acp{NIF} are trained on static scenes, requiring a re-training of the networks with each change of scene configuration or camera viewpoint \cite{fujieda2023neural}. 
Our neural bounding networks, in contrast, are learned on object-level, and hence can easily be rearranged in a scene without retraining, as we show in our experiments. 
Both our method and \ac{NIF} are inspired by neural fields, that have successfully modeled occupancy \cite{mescheder2019occupancy}, signed distance and surface distance \cite{park2019deepsdf,behera2023neural}.
For a comprehensive survey of recent works employing coordinate-based NNs, please see \cite{xie2022neural}.

In concurrent work, not specific to rendering, very simple primitives are fitted conservatively to polytopes \cite{hashimoto2023neural}, but unable to handle general shapes.
Neural concepts have been used to create bounding sphere hierarchies by Weller and colleagues \shortcite{weller2014massively}, which uses a neural-inspired optimizer, where the representation of the bounding itself remained classic spheres, while in this article we use non-linear functions.
Others \cite{zesch2022neural} attempt to optimize collision testing by replacing the test with a neural network.
While that work is similar to ours in the sense that it represents the bounding itself as a non-linear function, it does not strictly bound but simply fits the surface of the indicator with a \ac{MLP} under a common loss.
This is also applicable to higher-dimensional spaces (C-spaces) of, \eg, robot configurations \cite{cai2022active}.
Essentially, these methods train signed distance or occupancy  functions, but without any special considerations for the difference of \ac{FP} and \ac{FN}, which is at the heart of bounding.
We compare to such approaches and show that we can combine their advantages with the benefit of never missing an intersection.
More advanced, combinations of fields can be learned so as to not collide \cite{santesteban2022ulnef}, but again only by penalizing intersections, not by producing conservative results.
Other constraints such as 
eikonality \cite{atzmon2020sal},
Lipschitz continuity \cite{yariv2021volume}, or
indefinite integrals \cite{nsampi2023neural} can be incentivized similar to how we incentivize conservativeness.
Sharp and Jacobsen \cite{sharp2022spelunking} have proposed a method to query any trained implicit \ac{NN} over regions using interval arithmetic.
That is orthogonal to the question of training the \ac{NN} to bound a function conservatively, which we study here.

To ensure no \acp{FN}, we make use of asymmetric losses, which are typically applied with aims different from ours, such as reducing class imbalance \cite{ridnik2021asymmetric}, to become robust to noise \cite{zhou2021asymmetric}, to regularize a space \cite{liu2023learning} or, closer to graphics, to control bias and variance in \ac{MC} path tracing denoisers \cite{vogels2018denoising}.

\section{Our Approach}\label{sec:OurApproach}
This section will outline how we construct our networks and the asymmetric loss in order to achieve tight, conservative bounding (strictly zero false negatives) in arbitrary dimensions.

\subsection{Method}\label{sec:Method}
To achieve our task of conservatively bounding in \indicatorDimension-dimensional spaces, we seek to learn a \ac{NN} that classifies concave regions of space into inside and outside, while ensuring strictly no false negatives. 
Input to our algorithm is an \indicatorDimension-dimensional \emph{indicator} function $\indicator(\location)\in\R^\indicatorDimension\rightarrow\{0,1\}$ that returns 1 inside and on the surface of the object, and 0 everywhere else.
In 2D, the indicator could be visualized as a regular image grid, a voxel grid in 3D, an animated object in 4D, or a multi-dimensional state space of robot arm poses (or even another network, \eg, a neural density field) in higher dimensions. 
We assume we can evaluate the indicator function exactly and at arbitrary coordinates.
It further is not required to differentiate this function with respect to anything.

On top of the indicator, we define a \emph{query} function $\query(\region)\in\R^\queryDimension\rightarrow\{0,1\}$ that is 1 if the indicator returns 1 for at least one point in the region \region.
For point queries, the indicator and query are identical, \ie, $\query = \indicator$.
For extended queries, such as range queries, $\region$ would be a parameterization of a region, \eg, the two corners defining an \ac{AABB}.
While in lower dimensions the indicator and region could be converted into another indicator (akin to the morphological ``open'' operation on images \cite{dougherty1992introduction}), our method also supports queries on high-dimensional indicators that can only be sampled and not be stored in practice.

At the core of our approach is another function $\bounding(\region)\in\R^m\rightarrow\{0,1\}$, with learnable parameters \parameters, that is strictly 1 where \query is 1, but is allowed to also be 1 in other places (\ac{FP}). 
While traditional approaches use computational geometry to infer \parameters (e.g., via the repeated projection step in \acp{KDOP} or the simple min/max-operation in \acp{AABB}), we leverage the power of gradient-based optimization of neural networks to learn the most suitable non-linear \bounding.

The training objective \loss to approximate \query via \bounding is the combined cost of all \acp{FN} and \acp{FP} across a region \region:
\[
\loss(\parameters)
=
\int
\cost(\region)
\mathrm d
\region,
\quad
\cost(\region)=
\begin{cases}
\begin{alignedat}{3}
0
&\text{ if }
\query(\region)=0 &&\text{ and }
\bounding(\region)=0, \quad \textrm{\acs{TN}}\\
\fnCost
&\text{ if }
\query(\region)=1 &&\text{ and }
\bounding(\region)=0, \quad \textrm{\acs{FN}}\\
\fpCost
&\text{ if }
\query(\region)=0 &&\text{ and }
\bounding(\region)=1, \quad \textrm{\acs{FP}}\\
0
&\text{ if }
\query(\region)=1 &&\text{ and }
\bounding(\region)=1, \quad \textrm{\acs{TP}}\\
\end{alignedat}
\end{cases}
,
\]
where \fnCost is the cost for \acl{FN}, which needs to be $\fnCost=\infty$ to be conservative, and $\fpCost$ is the cost for a \acl{FP}, which we define to be 1.
The first and last clause are true positive and negative and incur no cost, while the second clause ensures conservativeness, and the third ensures that the bounding is tight.

However, it is not obvious how to proceed with a loss that can be infinite.
Moreover, $\loss$ is discontinuous in $\parameters$ and has zero gradients almost everywhere, as the observed loss values only change in the proximity of the surface of the bounded region and are constant everywhere else.
While $\loss$ is required to ensure conservativeness, its optimization is infeasible for the aforementioned reasons.

We hence employ two modifications to \loss in order to make it usable in practice. 
First, we suggest to replace the fixed constants \fnCost and \fpCost with variable values $\fnCost(\learningTime)$ and $\fpCost(\learningTime)$ that depend on the learning iteration (for details, see Suppl. Sec. 2.1).
This ensures that, in the limit, the cost of \acp{FN} is unbounded, so the solution will eventually become conservative. 
Second, in order to compute smooth gradients for our neural bounding network \bounding, we approximate the previous \loss via a variant of a weighted binary cross-entropy \nnloss: 
\begin{equation}
\nnloss(\parameters) = 
-\mathbb E_i
[
\fnCost(\learningTime)
\cdot
\supervision_i \log(\prediction_{i,\parameters})
+
\fpCost(\learningTime)
\cdot
(1-\supervision_i)\log(1-\prediction_{i,\parameters})
], 
\label{eq:loss}
\end{equation}
where $\prediction _\parameters = \bounding(\location)$ is the bounding network prediction with the current parameters \parameters for the current input \location, and the supervisory signal $\supervision = \indicator(\location)$ is the result of evaluating the indicator \indicator at the same location.
The pseudocode for our loss is shown in \refAlg{Method}.

\begin{algorithm}[t]
    \caption[]{Conservative loss.
    \indicator is assumed to distribute over arrays like \samples.}
    \begin{algorithmic}[1]
    \Procedure{Loss}{\parameters}
        \State $\region = \Call{Uniform}{\null}$
        \State $\samples = \Call{SampleRegion}{\query, \region}$
        \State $\supervision = \Call{Any}{\indicator(\samples)}$
        \State $\prediction = \bounding(\location)$
        \State \Return $
        -\fnCost\cdot\supervision\cdot\Call{log}{\prediction}
        -\fpCost\cdot(1-\supervision)\cdot\Call{log}{1-\prediction}$
    \EndProcedure
    \end{algorithmic}
    \label{alg:Method}
\end{algorithm}

\subsection{Neural Bounding Hierarchies}\label{sec:Hierarchies}

We can also stack our neural bounding volumes into hierarchies, similar to how classic bounding boxes are used to compute a \ac{BVH}, with the added benefit that our neural hierarchy's higher levels are again tightly and conservatively bounding the inner  levels.
Please note that while the \acp{NN} bound each node, the tree's structure needs to be supplied by the user.

\subsection{Neural Early-out}\label{sec:EarlyOut}
One advantage of bounding hierarchies is that testing can be interrupted early under some conditions, saving computation time.
However, a typical \ac{NN} can only be executed to the end to produce a result.
We will next demonstrate we can also adapt our neural bounding approach to enable early-out as follows.

The idea is to use additional conservative and negated intermediate losses instead of a single conservative loss on the end.
Consider the example of an \ac{MLP} with two layers, \layerOne and \layerTwo, and a non-linearity \nonlinearity, performing a conservative inside-test.
The common loss, for simplicty at a single point, is 
\begin{align}
\mathcal L_{Late}(\region)
&=
\mathtt{bce}
(
\bounding^{Late}(\region),
\query(\region)), \text{ where }\\
\bounding^\mathrm{Late}(\region)
&=
\nonlinearity
\left(
\layerTwo
\times(\nonlinearity(\mathsf \layerOne\times(\region|1))|1)
\right)
,
\end{align}
{\change{with \texttt{bce} denoting the binary cross-entropy and $(\cdot|1)$ the bias-trick}. We instead train the loss}
\begin{align}
\mathcal L_\mathrm{Early}(\region)
&=
\mathcal L_\mathrm{Late}(\region)+
\mathtt{bce}(\bounding^\mathrm{Early}(\region), 1-\query(\region)), \text{where}\\
\bounding^\mathrm{Early}(\region)
&=
\nonlinearity(\mathsf \layerThree\times(\region|1)),
\end{align}

\noindent with $\layer_3$ being a third layer of suitable size.
Doing so, the network has to produce two conservative results at the same time: the final one, as well as an early one, that is the opposite of the first one.
The final one requires executing a long network.
The early one is typically a much simpler network.
At test time, the early network is executed first.
If its output is negative, we can be certain no further testing is required and exit.
If it is positive, the second network also gets executed  to get the final answer.
This is  possible as the early-out is inverse conservative, enabled by our approach.

This concept can further be cascaded to include multiple intermediate exit points.
It starts to be most effective if the \ac{NN} gets complex.
Our results show it to be around 25\,\% faster than a naive \ac{NN} across a substantial range of tasks and dimensions and about twice as fast on the important case of point queries. 

\subsection{Implementation}\label{sec:Implementation}

While our approach is realized as a neural field, making it generally applicable and independent of any specific architecture, we have observed that certain architectural decisions impact the tightness of the bounding. We detail these choices in the following sections.

\paragraph*{Architecture}
In all cases, the input to our algorithm is the indicator function to be bounded, which is then sampled at \queryDimension-dimensional query locations that are used to fit the network with our asymmetrical loss from \refAlg{Method}.
The output of the network is a floating point number that represents occupancy, restricted to $(0,1)$ via a Sigmoid and then rounded to $\{0,1\}$. 
For our concrete implementation, our network is implemented as \ac{MLP} in order to deal with arbitrary-dimensional queries. 
The architecture details for all results shown in this paper are reported in Suppl. Tab. 3.
Sinusoidal activations are used in the hidden layers, and the output layer is activated by a Sigmoid function.
We have experimented with both residual- and skip-connections \cite{he2016deep, ronneberger2015u} as well as Batch-Normalization \cite{ioffe2015batch}, but found little improvement, presumably due to the shallow network depth. 
For some results, we use positional encodings (see Suppl. Sec. 3).

\paragraph*{Training}
We build and train our networks in PyTorch \cite{paszke2019pytorch} and use the standard layer initialization, which we have found to especially perform favorably in higher dimensions.  
We use the Adam optimizer \cite{kingma2014adam} with a learning rate of $1\times10^{-3}$ and implement $\fnCost(\learningTime)$ and $\fpCost(\learningTime)$ as a linear step-wise schedule that is incremented every \num{10000} training iterations (see Suppl. Sec. 2.1 for details).
We use a batch size of \num{200000} and early-stop the training as soon as \ac{FN} = 0 is reached and has been stable for the past six scheduling iterations. 
Depending on dimensionality and query complexity, this typically takes between 20 to 60 minutes on a modern workstation. 

Please note that the concept of ``epochs'' or train/test splits applies differently to generalization across a continuous space:
For learning and validation, we randomly sample this space, and for testing we do the same.
Our proposed method does not aim to learn generalization across objects, but a generalization of bounds across the hypercube of space, time, query type, and combinations thereof.
We would like to emphasize that this is the same task which classic bounding geometry performs, where a bounding box would also not generalize from a bunny to, \eg, a dragon.

\section{Evaluation}\label{sec:Evaluation}

The analysis of our results is structured around studying different bounding methods (\eg, boxes, spheres, \acp{KDOP}, etc., see \refSec{Methods}) on different tasks, which we define as different query types in varying dimension (\refSec{Tasks}).

\subsection{Methods}\label{sec:Methods}
We evaluate our approach's performance against different classic bounding primitives.
First, we compare to axis-aligned and non-axis-aligned (oriented) bounding boxes (\method{AABox} and \method{OBox}), followed by bounding by a \method{Sphere} and its anisotropically scaled counterparts, axis-aligned and oriented ellipsoids (\method{AAElli} and \method{OElli}), respectively, which all can be fit in closed form \cite{schneider2002geometric}.
Another widely-used bounding method we consider is \acp{KDOP},  method \method{kDOP}, implemented following \citet{ericson2004real}.
We set $k$, the number of planes, to $4\queryDimension$, scaling with the dimensionality.

\method{BVH} are a method common for spatial queries in graphics with the benefit of early interruption and not having to compare every node in the hierarchy to a query.
Hence, we include a baseline, using a linear builder \cite{lauterbach2009fast} with Morton-order and median splits.
Queries are processed in batches with persistent work queues \cite{aila2009understanding}.
Note that we use the \ac{BVH} to perform bounding queries, also when applied to rays, and not to trace rays, \ie, find the nearest hit position.

\method{OurNN} and \method{OurNNEarly} implement our neural bounding method, without and with early-out, respectively, as detailed in \refSec{Implementation}.

To demonstrate our method is not limited to classic neural fields, we include a neural grid method \cite{karnewar2022relu}, called \method{OurReLUField}, which does not scale to high dimensions, but is extremely fast to train.

Finally, the application of our proposed asymmetric loss is not restricted to neural networks, but can also be applied to other optimization problems.
We therefore explore replacing our network with a set of \ac{KDOP} planes which are then optimized with our asymmetric loss.
We call this method \method{OurkDOP}, which has the speed and memory usage of traditional k-DOPs but benefits from parameters found by modern gradient descent.

To verify the contribution of our asymmetric training loss, we study a variant  of our approach that uses a symmetric loss which weights both \acp{FP} and \acp{FN} with the same weight, \ie, $\fnCost = \fpCost = 1$.
Due to its similarity to the classic occupancy networks \cite{mescheder2019occupancy}, we call this method \method{OccNet}.
Please note that \method{OccNet} is not a method that can be deployed for bounding in practical scenarios, as it does not produce conservative results (\ie, the false-negative rate is not zero). 
We would like to re-emphasize that deploying non-conservative bounding method in graphics would lead to missing geometry, i.e., rays that do actually intersect a 3D object will wrongly test negative (see column \method{OccNet} in \refFig{3DResults}).
We hence only show qualitative, not quantitative results of this method.

Because of numerics, it is typical that our and other methods designed to be conservative on an (infinite) training sample set might be non-conservative on a hidden test set.
Consider fitting a single plane to classify space in two: the exact plane equation might not be float-representable.
The same is true for the parameters of a NN which are also floating-point.
To account for this at finite training time, we add an epsilon (see supplemental) to each methods\change{'} output.
\change{With our methods, we find fewer than 1 \ac{FN} in 100 million queries on hidden test sets, and this will decrease with more training.}

\begin{figure}[t]%
\centering\includegraphics*[width =\linewidth]{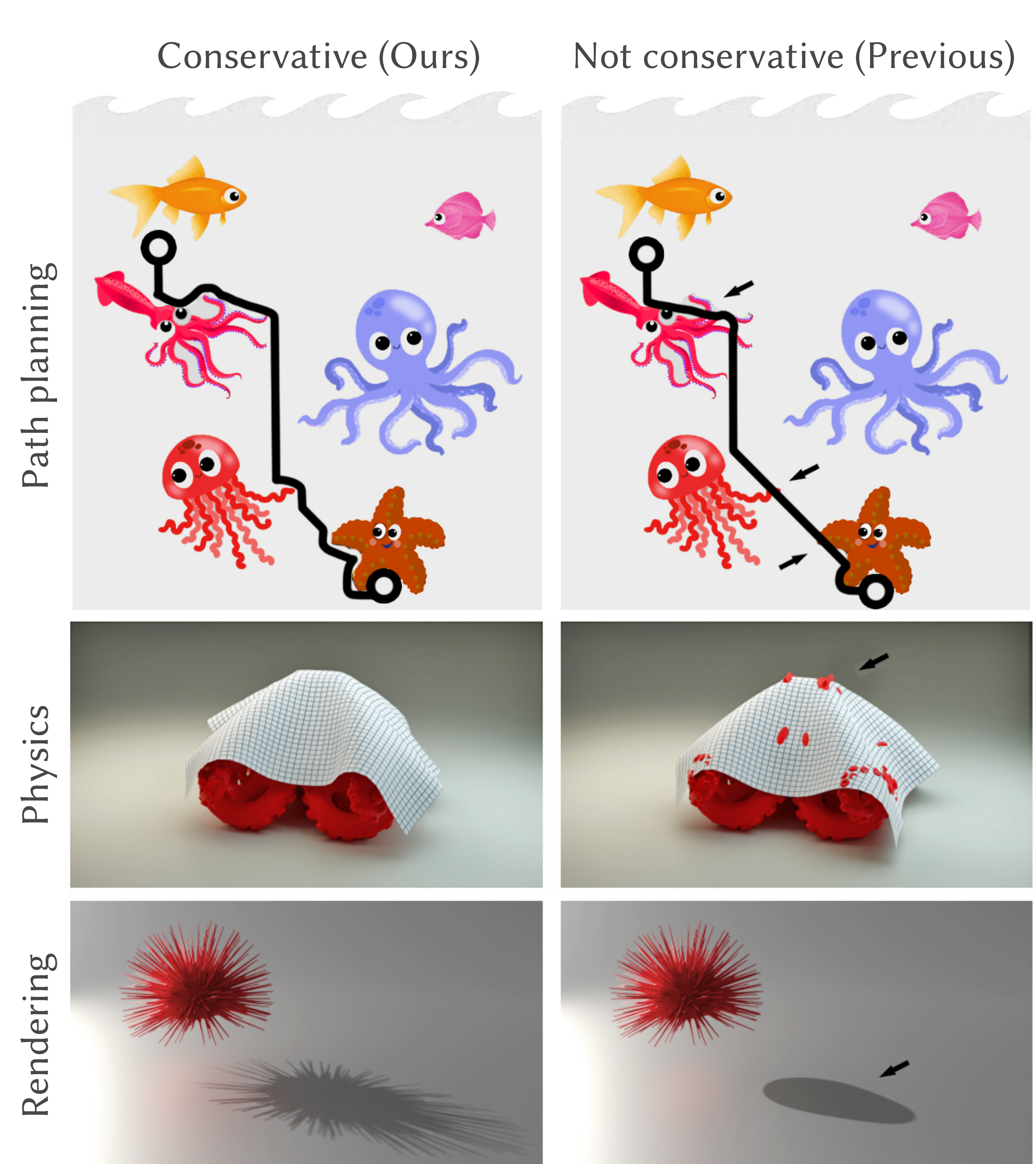}%
\vspace{-.15cm}%
\caption[]{{Motivating examples for conservative queries: we encode the world as a \ac{NN} trained with (left) and without (right) our proposed approach.
When used for 2D path-planning, our representation achieves a collision-free trajectory (left), whereas the non-conservative \ac{NN} collides (arrows).
Similar observations can be made when using the \ac{NN} as a proxy for 3D cloth collision leading to erroneous interpenetration in the middle row, or as a proxy to cull shadow rays where the stings are missing in the shadow.}}%
\vspace{-.15cm}%
\label{fig:Applications}%
\end{figure}%

\subsection{Tasks}\label{sec:Tasks}
In this analysis, a ``task'' combines two properties: the dimension of the indicator function (we study $\indicatorDimension=2$, $3$ and $4$) and the type of query (points, rays, planes and boxes), which builds up to eight-dimensional problems.
\change{\refFig{Applications} shows some examples of these queries.}

\myparagraph{Indicators}
For 2D data we use images of single natural objects in front of a white background where the alpha channel defines the indicator function.
We use 9 such images.
For 3D data, we use 9 voxelized shapes of popular test meshes, such as the Stanford Bunny and the Utah teapot.
\change{We show these objects in Suppl. Fig. 1}.
For 4D data, we study sets of animated objects: we load random shapes from our 3D data and create time-varying occupancy data by rotating them around their center.
We obtained 3 samples of this distribution and would like to emphasize that this is a strategy that favors the baselines: if object transformations were characterized by translational instead of rotational motions, the performance of the baseline approaches would deteriorate significantly. 
\method{AABox}, for instance, would have to bound the entire spatial extent between the initial and terminal object locations, thereby yielding an exceedingly high number of false-positive intersections.

\myparagraph{Query types}
Our query types are point-, ray-, plane- and box-queries. 
For all query types, the goal is to ask, given a point, ray, plane or box, does it intersect the object to be bounded?
Rays are parameterized as origin and direction vector, planes as normal and point $p_0$ on the plane surface, and boxes by their minimum and maximum corners. For every query region, the result is computed as \texttt{any()} of a sample of the indicator across the region.

\subsection{Metrics}\label{sec:Metrics}
We report results for the two relevant metrics which define the quality of a bounding approach: \emph{tightness} and execution \emph{speed}.

\myparagraph{Tightness}
In terms of tightness, \ac{FP} is the only figure of merit to study, as a high \ac{FP} rate will result in many unnecessary intersection tests. 
This is worsened if not working with hierarchies (or on their lowest level), as then every \ac{FP} leads to an expensive test against the entire bounded geometry, with often \change{millions} of triangles. 
We compute the \ac{FP} and \ac{FN} using \ac{MC} of one million values. 

Although our task is (space) classification, we do not employ classic metrics such as F1, precision and recall, 
as these capture the relation between \acp{FP} and \acp{FN}, which in our case is not relevant, since the \ac{FN} rate for all bounding methods must be strictly zero.

\myparagraph{Speed}
This is the speed of the bounding operation itself (\eg, evaluating the closed-form sphere intersection, or, for method \method{OurNN}, a network forward pass).
We report both query speed and ray throughput as the average number over \num{10000} independent runs with 10 million randomly sampled, forward-facing 3D rays.
For fairness, our methods and all baselines have been implemented as vectorized PyTorch code and make full use of GPU acceleration. 

\subsection{Results}\label{sec:Results}
\myparagraph{Quality} 
We show our quantitative results and the comparisons against the baselines in \refTbl{Results}. 
As is evident, our method \method{OurNN} consistently outperforms the other baselines by a large margin (up to 8$\times$ improvement on 4D point queries).
\method{BVH} performs second best across the tasks, roughly trailing ours by three or four times more \acp{FP}.
\begin{wrapfigure}{r}{0pt}%
 \vspace{-.2cm}%
    \includegraphics[width=0.35\columnwidth]{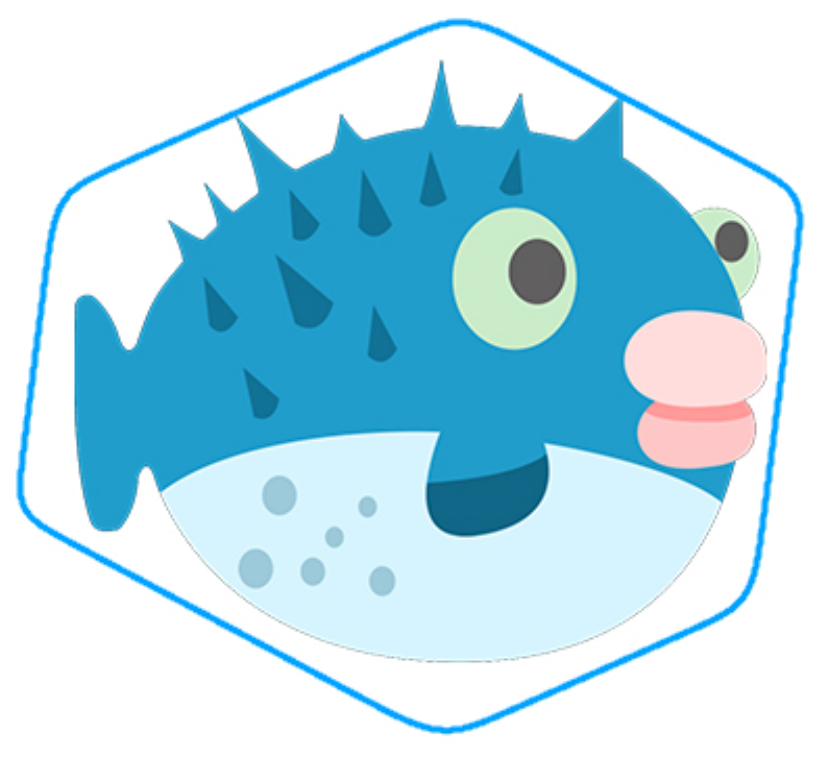}%
    \vspace{-.2cm}%
    \caption[]{Result for \method{OurkDOP}. \tiny{Image credit: 588ku, PNGTree.com}}%
    \label{fig:OurKDOP}%
    \vspace{-.2cm}%
\end{wrapfigure}%
The follow-up methods are on average \method{kDOP} and \method{OurkDOP}. 
Interestingly, using modern gradient-descent based construction for \method{OurkDOP} improves on \method{kDOP} results significantly. 
Unsurprisingly, all methods increase their \ac{FP} rate when going to higher dimension, a testimony to the increased query complexity. 
Notably, our method scales favorably with dimension and achieves acceptable \ac{FP} rates even at high dimensions (\eg, for 4D box queries). 
Moreover, especially in higher dimensions (\eg, 4D ray, plane, box), the baselines approach uniform performance, \eg, \method{AABox} vs \method{kDOP}, differing by only a few percentage points.
\method{BVH} does better here, getting closer to our performance, but still with a margin.
The relation of our variant \method{OurNNEarly} to \method{OurNN} is favorable: early-out creates slightly more \ac{FP} in some cases, but the gain is the win in test compute time studied later.
The other variant \method{OurReLUField} is performing worst of all neural methods, trains and tests extremely fast.
\refFig{Rank} visualizes the same data in the form of a rank plot, see figure caption for discussion.
Qualitative results for 2D and 3D are shown and discussed in \refFig{Teaser} and \refFig{3DResults}, respectively. 

In \refFig{OurKDOP} we show result for \method{OurkDOP} at $k=6$,
where the planes were found via stochastic gradient descent and our loss.

\begin{wrapfigure}{r}{0pt}%
 \vspace{-.2cm}%
    \includegraphics[width=0.35\columnwidth]{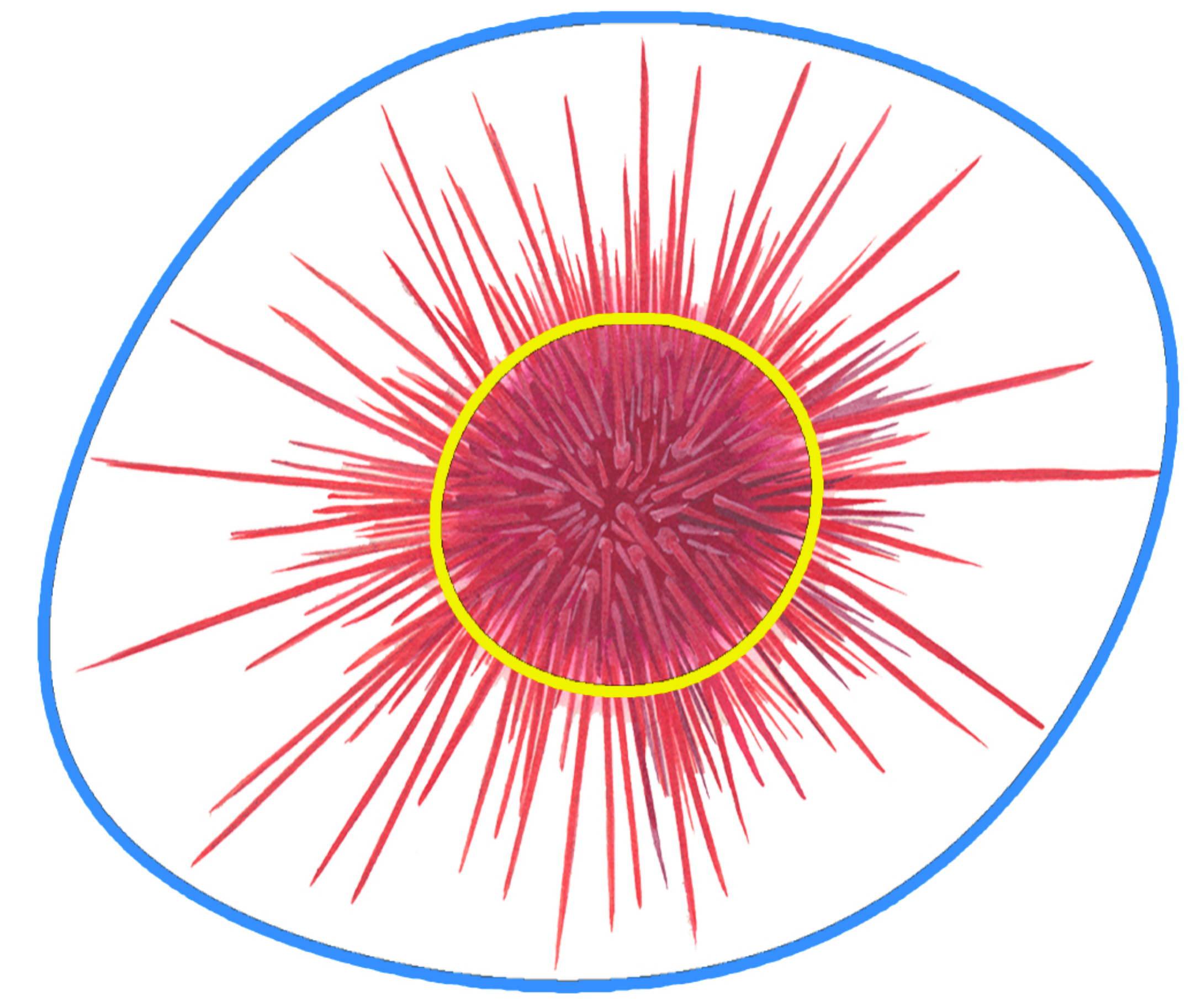}%
    \vspace{-.2cm}%
    \caption[]{Conservative boun\-ding  of never- (blue) and always-hit (yellow).}%
    \label{fig:Inverted}%
    \vspace{-.2cm}%
\end{wrapfigure}%
Ours performs better than the common heuristics \cite{ericson2004real}.
This indicates that our approach to finding bounding parameters can be superior to heuristics, even if the model itself is not neural. 

\begin{figure}[htb]%
\centering\includegraphics*[width = \linewidth]{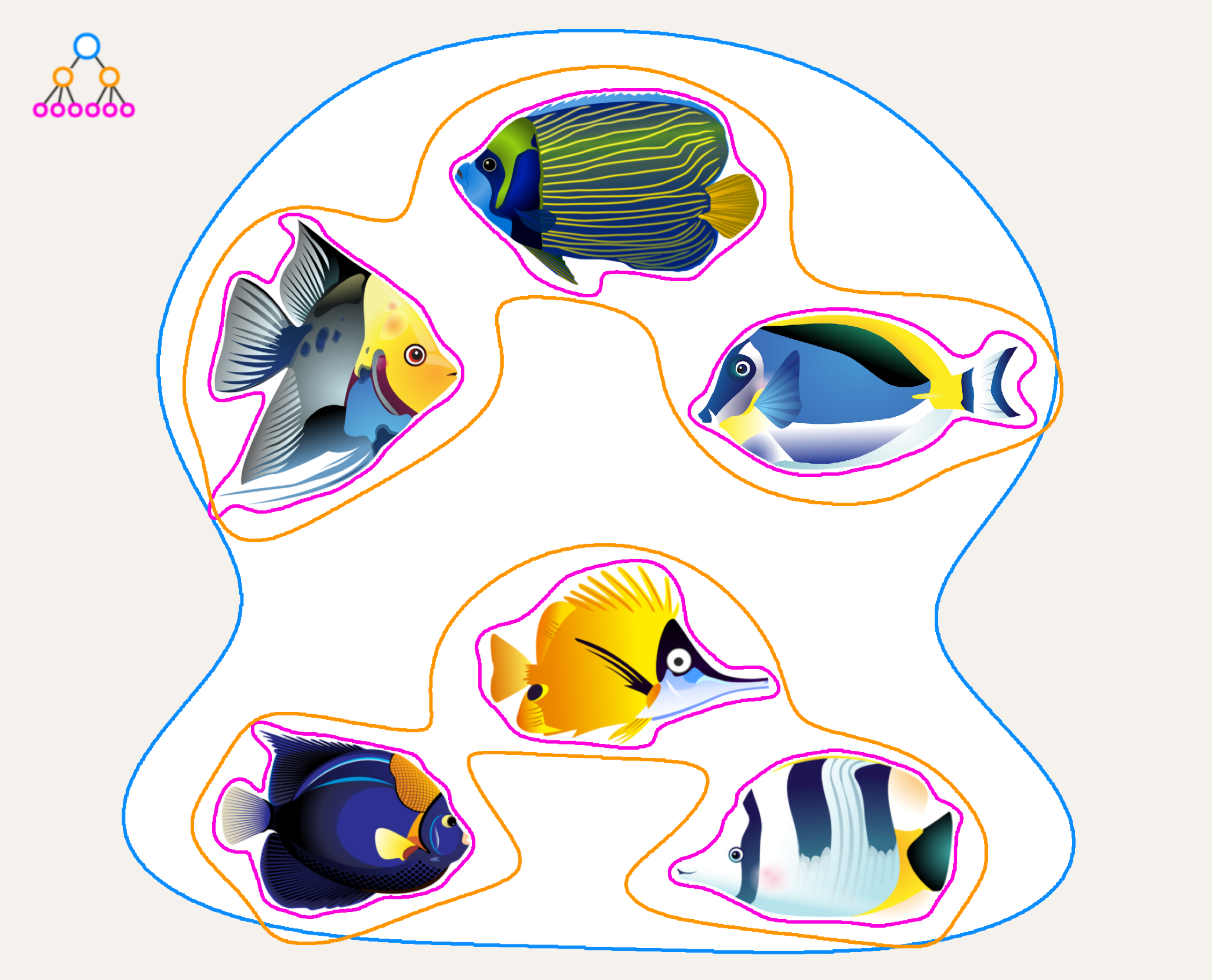}%
\vspace{-.15cm}%
\caption[]{A hierarchy of neural bounding networks.
The first level is show in blue, the two child nodes in yellow and the leafs in pink.
Note that it is not required for the higher-level bounding to bound the lower-level boundings.
It is only required to bound the indicator.}%
\vspace{-.15cm}%
\label{fig:BVH}%
\end{figure}

\begin{figure*}[htb]%
\centering\includegraphics[width = \linewidth]{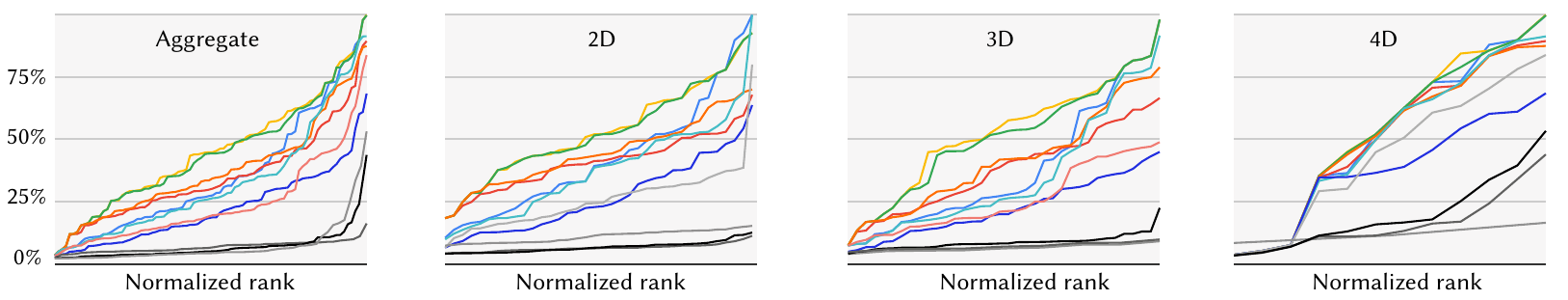}%
\vspace{-.2cm}%
\caption[]{\ac{FP} rates (vertical axis) for each test object, sorted in ascending order (horizontal axis). We show all methods (colored lines) across all dimensions (horizontal subplots).
While \refTbl{Results} reports average performance, this graphical representation reveals that our method \method{OurNN}, despite having a few higher negative examples, consistently yields the lowest \ac{FP} rates across all categories. The average of \method{OurkDOP} and \method{OurNNEarly} seem affected by some hard cases.}%
\vspace{-.2cm}%
\label{fig:Rank}%
\end{figure*}

We further show a result for an interesting variant of our method, that does not conservatively state which spatial location might be hit, but conservatively bounds which spatial locations certainly are hit.
This is achieved by flipping the asymmetry-weights \fnCost \, and \fpCost.
An example is seen in \refFig{Inverted} for 2D and \refFig{3DInverted} for 3D.
This is useful for a quick broad-phase test in collision: an object only needs to be tested if it is neither certainly out nor certainly-in.
\refFig{Progressive} shows our \method{OurNN}for \acp{NN} of different complexity.
\refFig{BVH} shows a bounding hierarchy on a school of 2D fish.

\myparagraph{Speed}
We quantify the speed of our bounding operations in \refTbl{Speed}.
While we obviously cannot match the speed of simpler bounding methods such as \method{Sphere} or \method{AABox}, we were surprised to find that our implementation of our neural bounding queries is not much slower than \method{kDOP} and some of the oriented bounding methods.  
Therefore, even if we nominally lag behind in this comparison (by a factor of min. 1.1$\times$, \method{OurNNEarly} vs \method{kDOP}, and max. 19.6$\times$, \method{OurNNEarly} vs. \method{Sphere}), we would argue that this is offset by the substantial reduction in \ac{FP} (on average $\times12$, see \refTbl{Results} and \refSec{Discussion}).  

\begin{table*}[ht]
    \centering
    \setlength{\tabcolsep}{6pt}
    \caption[]{False-positive rates for different bounding methods (rows) in different dimensions (columns) on different query types (subcolumns). 
    Lower means better, means fewer unnecessary tests.
    The best result per dimension and task is shown in bold.
    $^1$(Grids, like \method{OurReLUField} do not scale to higher dimensions.)}
    \begin{tabular}{r PPPP PPPP PPPP}
        \toprule
        &
        \multicolumn{4}{c}{2D}&
        \multicolumn{4}{c}{3D}&
        \multicolumn{4}{c}{4D}
        \\
        \cmidrule(lr){2-5}
        \cmidrule(lr){6-9}
        \cmidrule(lr){10-13}
        &
         \multicolumn{1}{c}{Point}&
         \multicolumn{1}{c}{Ray}&
         \multicolumn{1}{c}{Plane}&
         \multicolumn{1}{c}{Box}&
         \multicolumn{1}{c}{Point}&
         \multicolumn{1}{c}{Ray}&
         \multicolumn{1}{c}{Plane}&
         \multicolumn{1}{c}{Box}&
         \multicolumn{1}{c}{Point}&
         \multicolumn{1}{c}{Ray}&
         \multicolumn{1}{c}{Plane}&
         \multicolumn{1}{c}{Box}
        \\
        \midrule

\method{AABox} & 28.4 & 43.9 & 21.3 & 11.0 & 28.5 & 69.3 & 19.1 & 18.9 & 81.7 & 70.3 & 3.6 & 38.0 \\
\method{OBox} & 27.0 & 29.3 & 21.2 & 21.9 & 26.1 & 52.5 & 19.1 & 36.4 & 76.2 & 68.9 & 3.6 & 37.7 \\
\method{Sphere} & 40.2 & 43.9 & 21.3 & 27.3 & 56.7 & 69.3 & 19.1 & 47.3 & 84.5 & 70.4 & 3.6 & 40.6 \\
\method{AAElli} & 39.6 & 43.9 & 21.3 & 26.9 & 52.2 & 69.3 & 19.1 & 45.1 & 82.8 & 70.4 & 3.6 & 40.5 \\
\method{OElli} & 28.8 & 36.2 & 21.3 & 23.3 & 29.1 & 62.9 & 19.1 & 37.1 & 74.2 & 68.8 & 3.6 & 39.9 \\
\method{kDOP} & 28.4 & 33.4 & 21.2 & 8.9 & 22.0 & 65.6 & 19.1 & 17.4 & 75.3 & 70.0 & 3.6 & 36.4 \\
\method{BVH} & 10.4 & 28.9 & 15.3 & 8.2 & 10.0 & 34.4 & 18.4 & 14.6 & 39.8 & 61.7 & 3.6 & 35.2 \\
\hline
\method{OurkDOP} & 19.5 & 18.5 & 14.3 & 7.7 & 15.1 & 40.0 & 18.0 & 15.3 & 62.0 & 64.7 & 3.6 & 31.6 \\
\method{OurReLUField} & 5.0 & 6.4 & 5.7 & 3.3 & 3.3 & \notAvailable & \notAvailable & \notAvailable & \winner{10.0} & \notAvailable & \notAvailable & \notAvailable \\
\method{OurNN} & 3.2 & \winner{2.7} & \winner{1.2} & \winner{2.2} & 4.2 & \winner{7.6} & \winner{5.6} & \winner{5.0} & 10.9 & \winner{31.0} & 3.1 & \winner{11.3} \\
\method{OurNNEarly} & \winner{2.8} & 3.3 & 1.3 & 3.1 & \winner{2.9} & 10.1 & 5.7 & 6.5 & 11.5 & 38.8 & \winner{3.1} & 17.2 \\
    \bottomrule
\multicolumn{13}{@{}p{100pt}@{}}{\hspace*{-20pt}\includegraphics[width = 500pt]{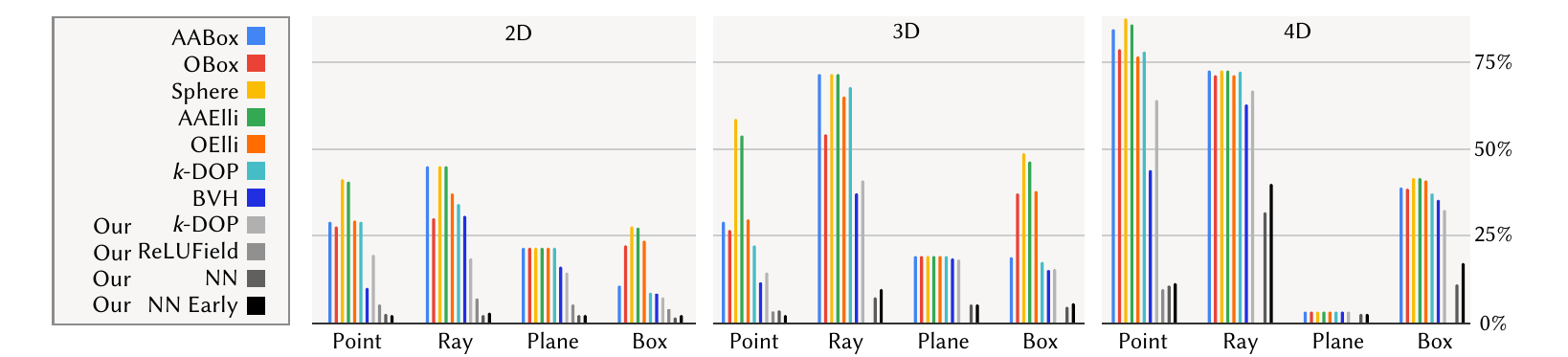}}\\
    \end{tabular}
    \label{tab:Results}
\end{table*}

\vspace{-.2cm}%

\begin{table}[t]
    \centering
    \caption[]{Speed and throughput of different methods.
     Throughput is in billions of rays per second. 
     The rightmost column shows the \change{factor by which} a geometry test has to be more expensive than querying the bounding method for our method (with early-out) to pay off, as per \refEq{Benefit}.}
    \begin{tabular}{lrrr}
        \toprule
        &
         \multicolumn{1}{c}{Speed (ms)}&
         \multicolumn{1}{c}{Throughput}&
         \multicolumn{1}{c}{Ratio} \\
        \midrule
        \method{AABox}& 2.58  & 3.88 & 78.78 \\ 
        \method{OBox} & 2.58  & 3.88 & 103.08 \\ 
        \method{Sphere} & 2.36  & 4.24 & 74.15 \\ 
        \method{AAElli} & 2.39  & 4.19 & 74.10 \\ 
        \method{OElli} & 2.39  & 4.18 & 83.07 \\ 
        \method{kDOP} & 41.27  & 0.24 & 9.06 \\ 
        \method{BVH} & \change{25.67} & \change{0.39} & \change{84.82} \\ 
        \hline
        \method{OurkDOP} & \change{11.84} & \change{0.84} & \change{114.92} \\ 
        \method{OurNN} & 70.82 & 0.14 &  --- \\ 
        \method{OurNNEarly} & 46.30  & 0.22 & --- \\ 
        \bottomrule
    \end{tabular}
    \label{tab:Speed}
\end{table}

\section{Discussion}\label{sec:Discussion}
Can a network that is slower than traditional bounding primitives be useful in practical graphics problems?
There are two supportive arguments we will discuss in the following: tightness and scalability.

\myparagraph{Tightness}
In order to evaluate the speed of a bounding method, one must additionally take into account the cost of a false-positive query, \ie, having to perform an intersection against the detailed, bounded geometry.
Assume a bounding method B can be queried in time $t_b$, and the competing bounding method A is five times slower, \ie, $t_a = 5 \, t_b$.
Assume further that A, in spite of being slower, produces significantly fewer false-positives ($p_a=0.1$) than B ($p_b=0.3$).
Finally, assume that performing tests with the actual detailed geometry needs time $t$, which is usually much larger than $t_a$ and $t_b$. 
The total time for \numTests tests, regardless of the method used, then is $\numTests t_i + p_i\cdot \numTests \cdot t$, of which the first term marks indispensable checks (as every ray must be checked against the bounding method), and the second term marks unnecessary checks due to false-positive bounding queries.
For method A to win, the following must hold:
\begin{align}
\numTests \cdot t_a + \numTests p_a\cdot t&< \numTests \cdot t_b + \numTests p_b \cdot t\nonumber\\
t_a + p_a\cdot t &< t_b + p_b\cdot t\nonumber\\
t_a + p_a\cdot t - p_b\cdot t &< t_b\nonumber\\
p_a\cdot t - p_b\cdot t &< t_b - t_a\nonumber\\
t (p_a - p_b) &< t_b - t_a\nonumber\\
\textrm{and hence if } (p_a - p_b) > 0: \quad t &< (t_b - t_a) / (p_a - p_b)\nonumber \\ 
\label{eq:Benefit}
\textrm{or otherwise if } (p_a - p_b) < 0: \quad t &> (t_b - t_a) / (p_a - p_b).
\end{align}
For the aforementioned example values, this produces
\[
t > (t_b - 5\,t_b) / (0.1 - 0.3) = 20\,t_b,
\]

\noindent
which means that method A is to be preferred if tests against the actual bounded geometry are at least 20 times as expensive as the bounding query. 
As traditional triangle meshes often have \change{millions} of triangles, this is easily achieved by our approach.

We quantify the factor by which a 3D ray-geometry test has to be more expensive than the bounding test in the right column of \refTbl{Speed} and see that our method is to be preferred when the geometry test is as little as 9$\times$ more expensive (for \method{kDOP}, avg. over all methods is 70.2), which certainly is achieved in most real-world applications.

\begin{wrapfigure}{r}{0pt}%
 \vspace{-.2cm}%
    \includegraphics[width=0.57\columnwidth]{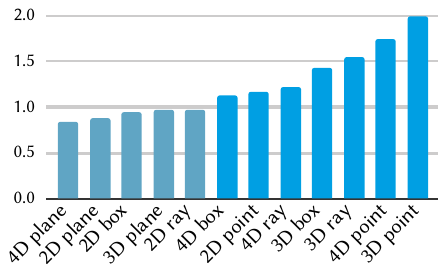}%
    \vspace{-.2cm}%
    \caption[]{Early-out speed (see text).}%
    \label{fig:EarlyOutPlot}%
    \vspace{-.2cm}%
\end{wrapfigure}%
\myparagraph{Early-out}
In \refFig{EarlyOutPlot} we analyze the effectiveness of  \method{OurNNEarly}, relative to \method{OurNN} in more detail.
The plot shows the ratio of both method's runtime, across all our tasks and dimensions, sorted from low to high.
There are four cases, mostly plane-type, where the speed gain is marginally below one. 
Otherwise, it is positive, including the most important case of 3D point queries being twice as fast.
The mean gain is 24.3\,\%.

\begin{wrapfigure}[10]{r}{0pt}%
 \vspace{-.2cm}%
    \includegraphics[width=0.55\columnwidth]{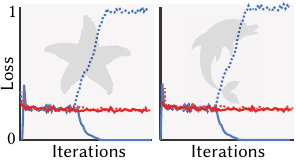}%
    \vspace{-.2cm}%
    \caption[]{Convergence (see text).}%
    \label{fig:ConvergencePlots}%
    \vspace{-.2cm}%
\end{wrapfigure}%
\myparagraph{Convergence}
In \refFig{ConvergencePlots}, we analyze the convergence for two scenes.
Each plot horizontally shows iterations and loss vertically.
Our method is shown in blue, the common loss in red.
FN in solid, FP in dotted. 
We see that ours, when the loss is ramped, reduces the FN to zero, at the expense of growing FPs.
\refFig{ReLUFields} shows that our approach can equally well be applied to non-\ac{MLP} architectures.
\change{Overall, our method incurs a training overhead.
Representations that train slowly remain slow, but if training is fast, it remains fast with our method (see \refFig{ReLUFields}).}

\myparagraph{Scalability}
To demonstrate scalability to complex high-dimensional spaces, we finally bound an entire generative model itself in \refFig{VAE}.
We use a pre-trained \ac{VAE} of MNIST digits as the indicator.
This is a 12-dimensional space of 10 VAE latent dimensions and the 2D pixel coordinate.
Note that a pre-trained VAE is a deterministic 12-D indicator function, independent of the question of whether it was trained non-deterministically.
This allows to predict which pixels will belong to a digit without even running the generative model.
While this is a toy problem, in more engineered applications it would, \eg, allow conservative ray intersection with complex 3D models that have not even yet been generated.

\myparagraph{Limitations}  
However, our approach also comes with certain limitations, which we will highlight here to inspire future work: 

\change{Foremost, our method neither provides an algorithmic guarantee to have no \acp{FN}, nor is it a new bounding primitive.
Instead, we heuristically incentivize conservativeness via our proposed loss (\refEq{loss}).
Future investigations will have to explore algorithmic conservativeness and other forms of hard constraints in \ac{NN} training.}

\change{Our method does not generalize across objects, but --as is common for neural fields-- is trained per-object}. 
The training takes significantly longer than inferring traditional bounding primitives, many of which have closed-form solutions, which
\change{has practical implications for animated scenes.}
While this can easily be outsourced to a pre-processing stage, there is potential in maximizing training speed by either incorporating the fully-fused \ac{NN} architecture proposed by \cite{muller2022instant} or by meta-learning \cite{sitzmann2020metasdf, fischer2022metappearance, tancik2021learned} a space of bounding networks, which would be especially useful for bounding similar geometry that only slightly differs in shape or pose.

\change{Furthermore, our method's intersection speed is not yet competitive with methods from 3D ray tracing, which benefit from years of optimization research, such as ray-box or \ac{BVH} intersections.}

\change{Additionally, sampling the indicator only works if object- and query-dimensions align. 
An example is sampling points in 2D space but testing against a 1D line, which will not work, as the probability of sampling a point on the line is zero. 
This can be alleviated by re-parameterizing the sampler to sample directly on the line instead.}

Finally, as we overfit a network per shape, our approach requires additional storage in the order of magnitude of the network parameters (\eg, 2,801 parameters for a 3D point query network), which naturally is larger than for traditional baselines (see \refTbl{Speed}). 
However, storage is cheaper than compute, and moreover, applications which use bounding geometry usually handle \change{large meshes with many triangles}, leading us to believe that our modest storage requirement for the network parameters is negligible in comparison.

\section{Conclusion}\label{sec:Conclusion}
In future work, the idea of asymmetric losses might be applicable to other neural primitives like hashing.
In a similar vein, distance fields could be trained to maybe underestimate, but never overestimate distance so as to aid sphere tracing.
It is further conceivable to bound not only geometry but also other quantities like radiance fields or their statistics for image synthesis.

\begin{wrapfigure}{r}{0pt}%
\centering\includegraphics[width=.45\linewidth]{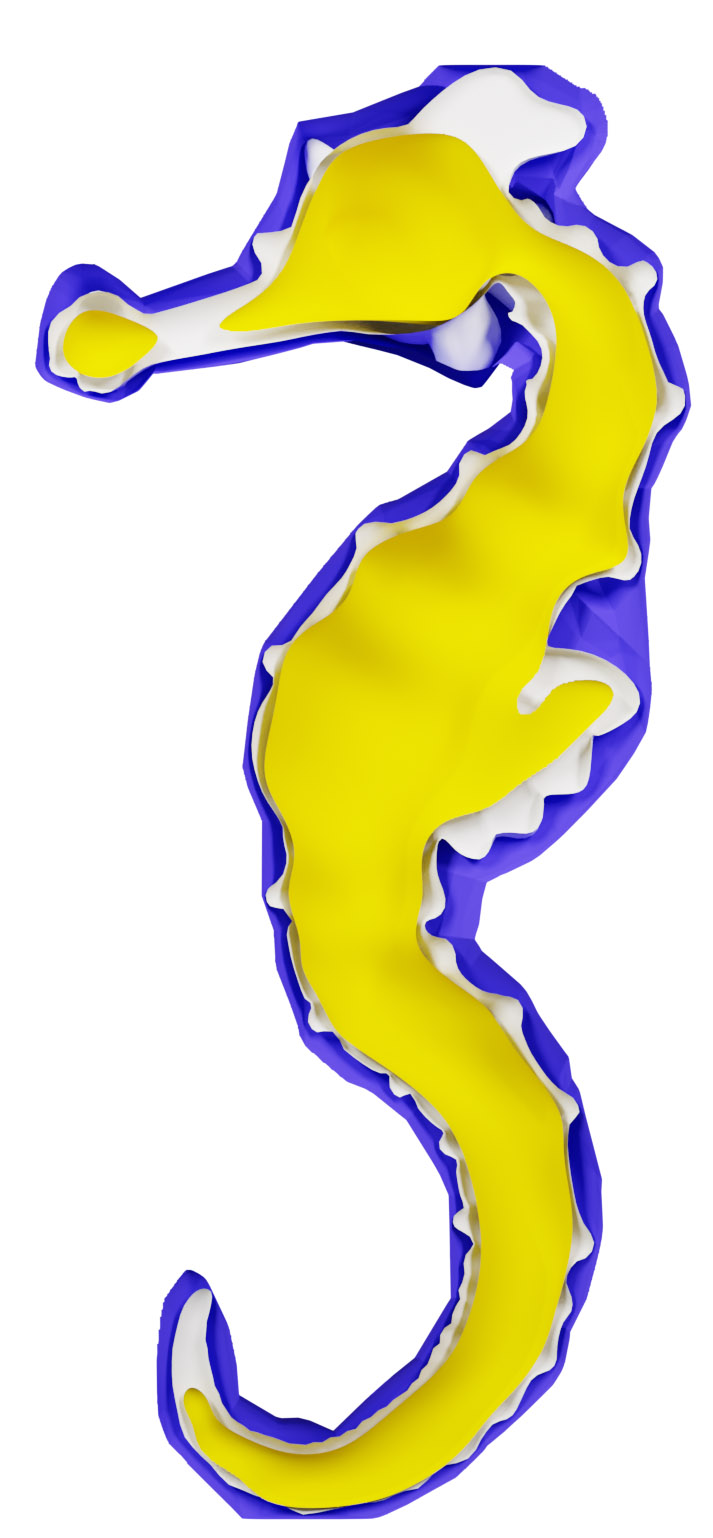}%
\vspace{-.2cm}%
\caption[]{3D-bounding no-hit (blue) and hit (yellow).}
\label{fig:3DInverted}
\end{wrapfigure}

Our results are only almost always conservative, as they involve two stages of sampling, that, in expectation, will be conservative, but we lack any proof under what conditions the probability of being truly conservative is how high.
The step of sampling the query region could be replaced by an unbiased (and potentially closed-form) one; we only choose sampling here as it works on any indicator in any dimension,
However, the loss is still an empirical loss, and there is a nonzero chance that a tiny part of the indicator would remain unattended with finitely many samples.
In practice, our results show dozens of tasks with many instances, each tested with hundred-thousands of samples with no \ac{FN} result.

Using our technique, bounding in graphics can benefit from many of the exciting recent innovations around \acp{NN} such as improved architectures, advanced training, or dedicated neural hardware.

\bibliographystyle{ACM-Reference-Format}
\bibliography{paper}

\appendix
\newpage

\begin{figure*}[htb]%
\centering\includegraphics[width =\linewidth]{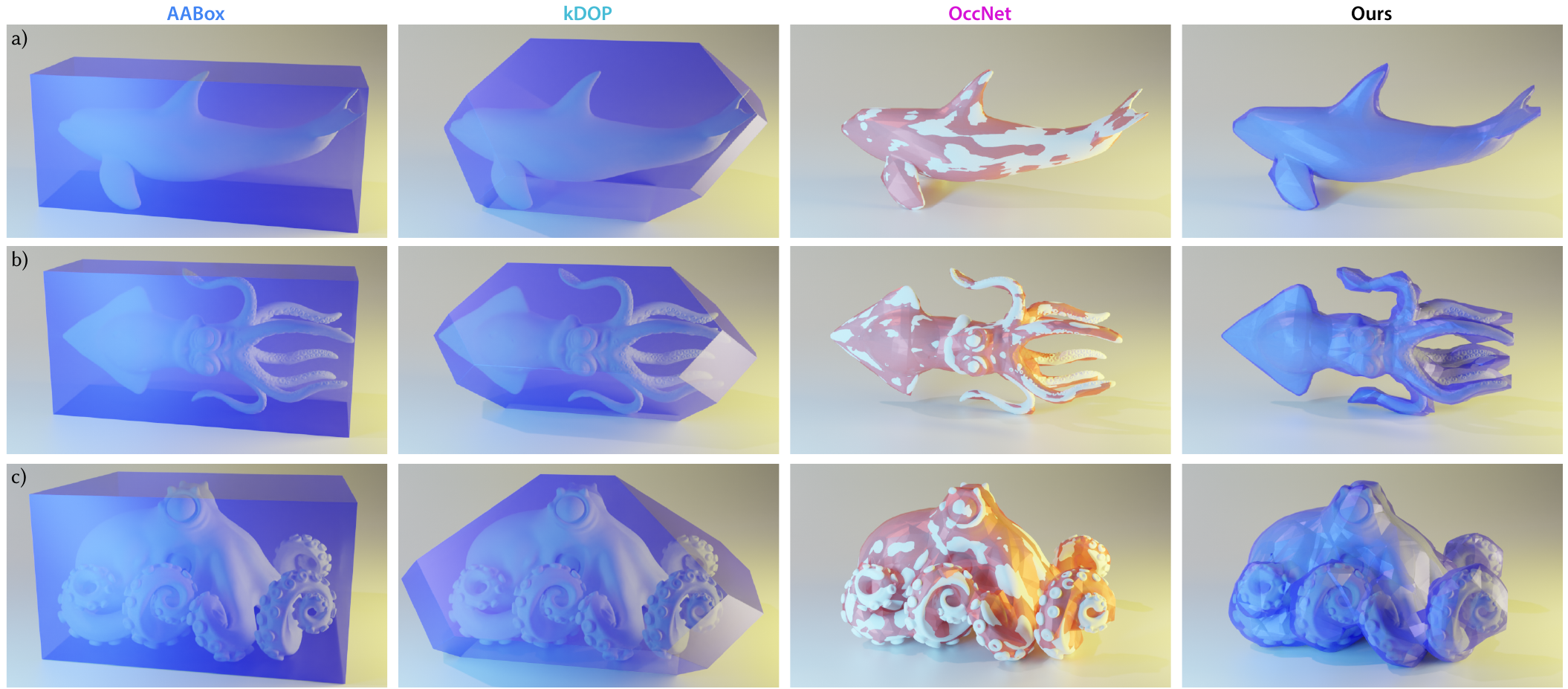}%
\vspace{-.15cm}%
\caption[]{Comparison of methods (columns) for different 3D shapes (rows).
\method{AABox} and \method{kDOP} both bound the objects conservatively, but create false positives, whereas \method{OccNet} bounds overtight and produces false negatives (red).
Our approach is both tight and conservative.}%
\vspace{-.15cm}%
\label{fig:3DResults}%
\end{figure*}%

\begin{figure*}[htb]%
\centering\includegraphics[width = \linewidth]{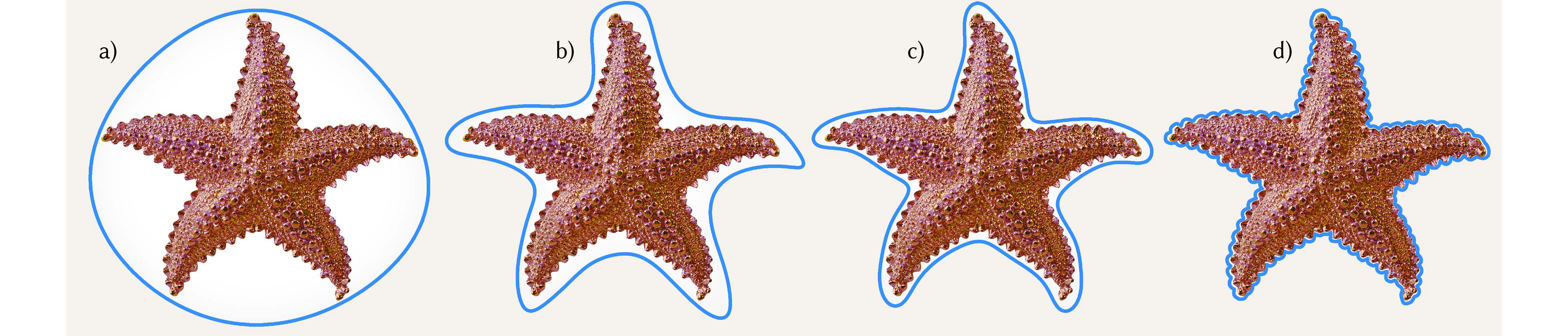}%
\vspace{-.15cm}%
\caption[]{{Our learned bounding volumes with increasingly complex neural networks from left to right.
A simple network, left, fits the shape only roughly, but still conservative.
With increasing expressiveness towards the right, the fit gets more accurate and recovers concave regions, and later, the curvature of the starfish's arms.
For a complex \ac{NN} it becomes almost pixel-exact, but even then, small approximation errors are made but never violating the bounding requirement (zero FN).
This indicates we can achieve conservativeness independent of complexity.}}%
\vspace{-.15cm}%
\label{fig:Progressive}%
\end{figure*}%

\begin{figure}[htb]%
\centering\includegraphics[width = \linewidth]{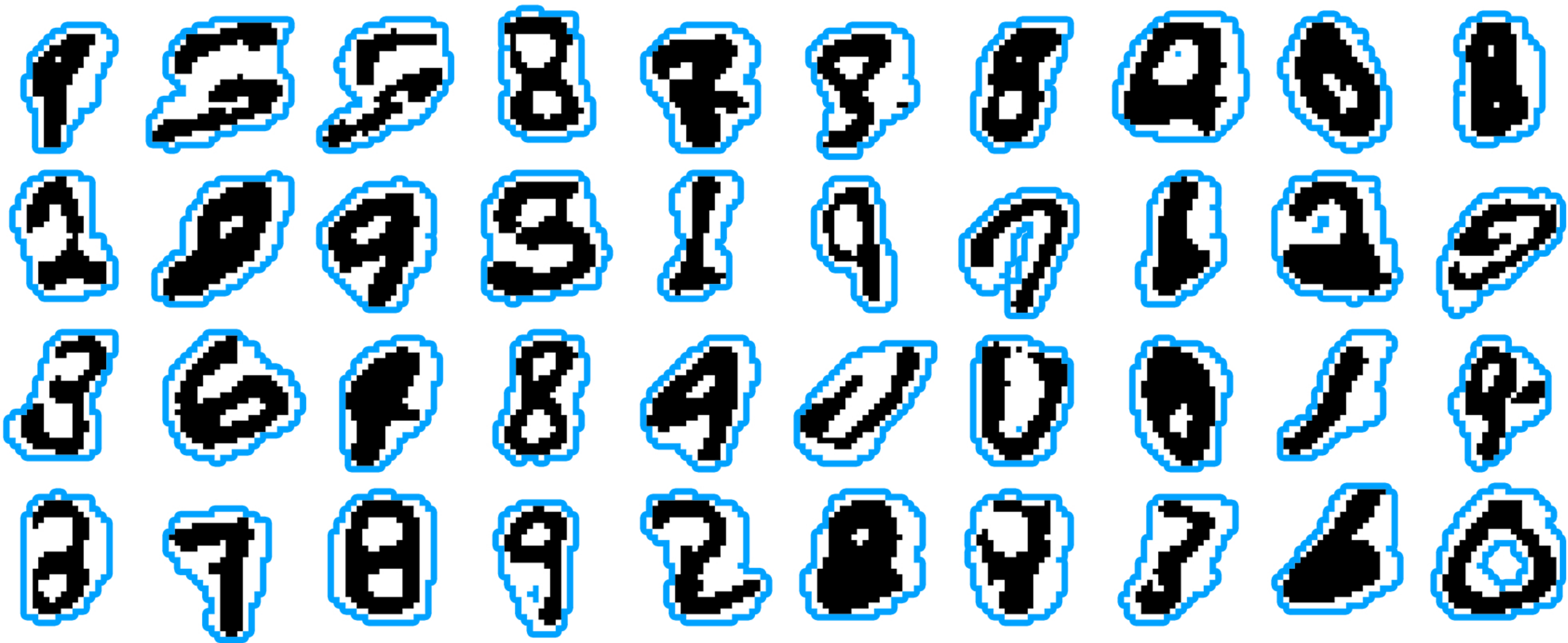}%
\vspace{-.15cm}%
\caption[]{{Results of applying our method to a \ac{VAE} that generates MNIST digits, here shown for random latent codes. The blue borders are our bounds.}}%
\vspace{-.15cm}%
\label{fig:VAE}%
\end{figure}%

\begin{figure}[htb]%
\centering\includegraphics[width = \linewidth]{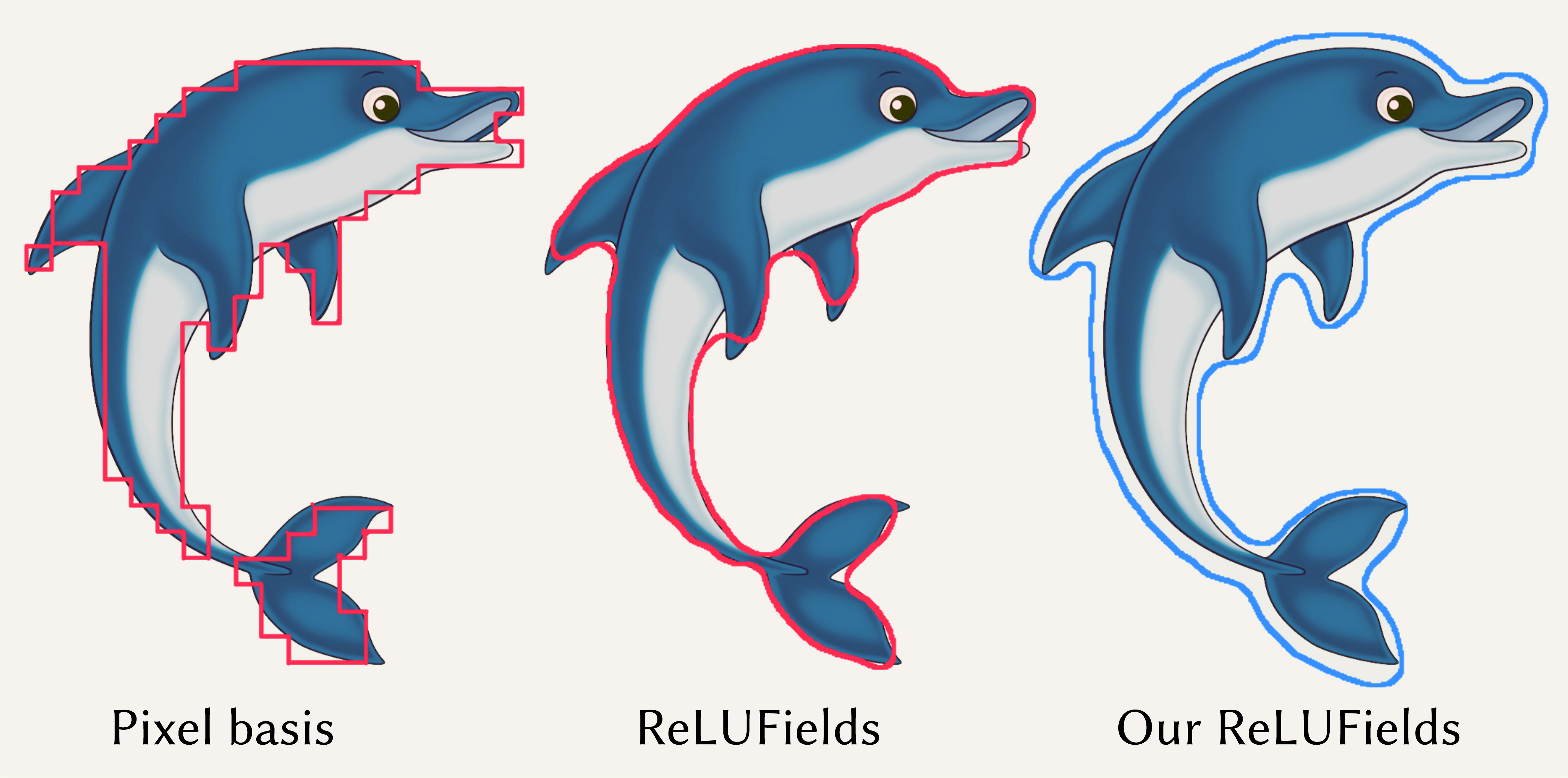}%
\vspace{-.15cm}%
\caption[]{{Application of our training to ReLUFields \cite{karnewar2022relu}, an example of a fast-trainable architecture.
Left shows a pixel basis result, middle a ReLUField of the same resolution and left our result.}}%
\vspace{-.15cm}%
\label{fig:ReLUFields}%
\end{figure}%

\newpage

\end{document}


\title[Neural Bounding: Supplemental]{Neural Bounding: Supplemental}

\author{Stephanie Wenxin Liu}
\affiliation{%
	\institution{Birkbeck, University of London}
	\country{United Kingdom}
}
\email{wenxin.liu.cs@gmail.com}

\author{Michael Fischer}
\affiliation{%
	\institution{University College London}
	\country{United Kingdom}
}
\email{m.fischer@cs.ucl.ac.uk}

\author{Paul D. Yoo}
\affiliation{%
	\institution{Birkbeck, University of London}
	\country{United Kingdom}
}
\email{p.yoo@bbk.ac.uk}

\author{Tobias Ritschel}
\affiliation{%
	\institution{University College London}
	\country{United Kingdom}
}
\email{t.ritschel@ucl.ac.uk}

\maketitle

\makeatletter
\newcommand{\nndefine}[1]{%
  \comma@parse{#1}{\nnentry}1%
}
\newcommand*{\nnentry}[1]{%
  #1$\times$%
}
\makeatother

\mysection{Introduction}{Introduction}
This document provides details on training used (\refSec{Training}), methods investigated (\refSec{Methods}), data tested (\refSec{Data}), speed observed (\refSec{Speed}) and an analysis of complexity (\refSec{Complexity}) of our method. 

\mysection{Training}{Training}

Our training uses a combination of scheduling the weights of different loss terms (\refSec{Schedule}) and regularization (\refSec{Regularization}).

\mysubsection{Weight schedule}{Schedule}
\change{
We provide more details on the $\alpha(t)$ and $\beta(t)$ schedules mentioned in the main document. 

$\alpha(t)$ is a linear schedule that depends on the indicator dimension. For 2D, $\alpha(t)$ decreases by 1/20, 1/40, 1/60 etc.; for 3D, $\alpha(t)$ decreases by 1/100, 1/200, 1/300, etc.; for 4D, $\alpha(t)$ decreases by 1/200, 1/400, 1/600, etc., every 10,000 iterations if not 0 \acp{FN}. 
We observe that we need a greater degree of asymmetry to achieve conservativeness in higher dimensional space, because as the dimensionality increases, the number of samples required to adequately cover the space also increases. Therefore to effectively eliminate \acp{FN} in a greater number of samples requires a bigger degree of asymmetry.

$\beta(t)$ is a linear schedule that increases by  $0.2$ every 10,000 iterations if not 0 \acp{FN}. This is just a tweak in order to speed up training - decreasing $\alpha(t)$ alone already achieves conservativeness, as conservativeness only depends on the degree of asymmetry. 
}

\mysubsection{Regularization}{Regularization}
\change{
We use L1 and L2 regularizations in our methods, lambda ranging from 0 to $5e-7$ increasing in a linear schedule.
}

\mysection{Methods}{Methods}
To make methods most comparable, we handle numerics fairly (\refSec{Numerics}) and use the same input parameterization (\refSec{Parameterization}). 

\mysubsection{Numerics}{Numerics}
\change{
An epsilon value of $1e-5$ is applied to the output of all methods to account for numerics.
}

\mysubsection{Parameterization}{Parameterization}
\change{
We encode rays as pairs of Cartesian $n$D start- and direction-vectors; boxes as pairs of Cartesian $n$D corners and planes as $n$D $p_0$ and normal.
}

\mysubsection{Our architecture}{OurNNArchitecture}
We detail the architectures used to produce the main paper figures in \ref{tab:Architecture}.
``Main Tab. $i$'' refers to table $i$ in the main document. 

Further, details for other methods are summarized in \refTbl{OurkDOPArchitecture} (\method{kDOP}),
\refTbl{OurReLUFieldArchitecture} (\method{OurReLUField}),
\refTbl{OurNNArchitecture} (\method{OurNN})
and \refTbl{OurNNEarlyArchitecture} (\method{OurNNEarly}).

\change{
For the positional encoding mentioned in the main document, we use the well-known approach presented in \cite{mildenhall2021nerf} to encode the network's input (details in \refTbl{Architecture}). 
We use an encoding depth of 8, which results in the input being 18-dimensional (2D point * 8 encodings, plus the 2D input).  
}

For point queries, the grid resolution is the same as our data size.
For ray, plane and box queries, since the input is 2$\times$ dimensionality, it is the data size doubled.

\begin{table}
\centering
\caption{\method{OurkDOP} architectures, where '$k$' is the number of directions.}
\label{tab:OurkDOPArchitecture}
\begin{tabularx}{\linewidth}{lrcrr}
\toprule
\multicolumn1c{Result}&
\multicolumn1c{Ind.}&
\multicolumn1c{Query}&
\multicolumn1c{$k$} \\ 
\midrule
Main Tab. 1 & 2D, 3D, 4D & Point, Ray, Plane, Box & $4$\\
\bottomrule
\end{tabularx}
\vspace{0.3cm}

\caption{\method{OurReLUField} resolutions.}
\label{tab:OurReLUFieldArchitecture}
\begin{tabularx}{\linewidth}{lrcrr}
\toprule
\multicolumn1c{Result}&
\multicolumn1c{Ind.}&
\multicolumn1c{Query}&
\multicolumn1c{Resolution} \\
\midrule
Main Tab. 1 & 2D & Point & $32 \times 32$\\
Main Tab. 1 & 2D & Ray, Plane, Box & $32 \times 32 \times 32 \times 32$\\
Main Tab. 1 & 3D & Point & $32 \times 32 \times 32 \times 32$\\
Main Tab. 1 & 4D & Point & $10 \times 32 \times 32 \times 32$\\
\bottomrule
\end{tabularx}
\vspace{0.3cm}

\caption{\method{OurNN} architectures.}
\label{tab:OurNNArchitecture}
\begin{tabular}{lrcrr}
\toprule
\multicolumn1c{Result}&
\multicolumn1c{Ind.}&
\multicolumn1c{Query}&
\multicolumn1c{Network} \\ 
\midrule
Main Tab. 1, Main Fig. 6 & 2D & Point & \nndefine{2, 25, 25}\\
Main Tab. 1, Main Fig. 6 & 2D & Ray, Plane, Box & \nndefine{4, 25, 25} \\
Main Tab. 1, Main Fig. 6 & 3D & Point & \nndefine{3, 50, 50}\\
Main Tab. 1, Main Fig. 6 & 3D & Ray, Plane, Box & \nndefine{6, 50, 50} \\
Main Tab. 1, Main Fig. 6 & 4D & Point & \nndefine{4, 75, 75} \\
Main Tab. 1, Main Fig. 6 & 4D & Ray, Plane, Box & \nndefine{8, 75, 75}\\
\bottomrule
\end{tabular}
\vspace{0.3cm}

\caption{\method{OurNNEarly} architectures.}
\label{tab:OurNNEarlyArchitecture}
\begin{tabular}{lrcrrr}
\toprule
\multicolumn1c{Result}&
\multicolumn1c{Ind.}&
\multicolumn1c{Query}&
\multicolumn1c{Input}&
\multicolumn1c{Exit1}&
\multicolumn1c{Exit2} \\ 
\midrule
Main Tab. 1 & 2D & Point & \nndefine{2} & \nndefine{25} & \nndefine{25, 25}\\
Main Tab. 1 & 2D & Ray, Plane, Box & \nndefine{4} & \nndefine{25} & \nndefine{25, 25}\\
Main Tab. 1 & 3D & Point & \nndefine{3} & \nndefine{50} & \nndefine{50, 50}\\
Main Tab. 1 & 3D & Ray, Plane, Box & \nndefine{6} & \nndefine{50} & \nndefine{50, 50}\\
Main Tab. 1 & 4D & Point & \nndefine{4} & \nndefine{75} & \nndefine{75, 75}\\
Main Tab. 1 & 4D & Ray, Plane, Box & \nndefine{8} & \nndefine{75} & \nndefine{75, 75}\\
\bottomrule
\end{tabular}

\end{table}

\begin{table*}[h]
\caption{Architecture details for all results shown in this paper. \(a \in \{10, 16, 25, 50, 75, 125\}, b \in \{25, 50, 64, 75, 125, 256\}, c \in \{75, 125, 256, 512, 768, 1024\}\)}
\label{tab:Architecture}
\centering
\begin{tabular}{lrcrr}
\toprule
\multicolumn1c{Result}&
\multicolumn1c{Ind.}&
\multicolumn1c{Query}&
\multicolumn1c{Network}&
\multicolumn1c{PE}\\
\midrule
Main Fig. 1 & 2D & Point & \nndefine{2, 20, 10, 5} & 0 \\
Main Tab. 1, Main Fig. 6 & all & all & \refTbl{OurkDOPArchitecture}, \refTbl{OurReLUFieldArchitecture}, \refTbl{OurNNEarlyArchitecture},  \refTbl{OurNNArchitecture} & 0\\
Main Tab. 2 & 3D & Ray & \nndefine{6, 50, 50} & 0 \\
Main Fig. 10, a, b & 3D & Point & \nndefine{3, 100, 100, 100} & 0 \\
Main Fig. 10, c & 3D & Point & \nndefine{3, 50, 50} & 0 \\
Main Fig. 4 & 2D & Point & \nndefine{2, 10} & 0 \\
Main Fig. 11, a & 2D & Point & \nndefine{2, 10, 10} & 0 \\
Main Fig. 11, b & 2D & Point & \nndefine{2, 10, 10, 5} & 0 \\
Main Fig. 11, c & 2D & Point & \nndefine{2, 25, 25, 5} & 0 \\
Main Fig. 11, d & 2D & Point & \nndefine{2, 128, 128, 128} & 18 \\
Fig. 2, 2D & 2D & Point & \nndefine{2, $a$, $a$} & 0 \\
Fig. 2, 3D & 3D & Point & \nndefine{3, $b$, $b$} & 0 \\
Fig. 2, 4D & 4D & Point & \nndefine{4, $c$, $c$} & 0 \\
Main Fig. 5 & 2D & Point & {$2\!\times\!10\!\times\!10\!\times\!10\!\times\!1$
 3 exits} & 0 \\
Main Fig. 12 & 12D & Point & \nndefine{12, 25, 25} & 0 \\
Main Fig. 9, hit & 3D & Point & \nndefine{3, 40, 40} & 0 \\
Main Fig. 9, no-hit & 3D & Point & \nndefine{3, 100, 100, 100} & 0 \\
\bottomrule
\end{tabular}
\end{table*}

\mycfigure{2D_3D_objects}{\change{The 2D  and 3D objects used for evaluation in the main manuscript. The 3D shapes are from ShapeNet \cite{chang2015shapenet} and the Stanford Computer Graphics Laboratory. We use the 2D images at resolution $32^2$ and 3D shapes in voxelized form at $32^3$ resolution. For the 4D sequences, the Stanford Bunny, Stanford Asian Dragon, and Utah Teapot are each rotated clockwise around their x-axis, completing a full rotation in 10 time steps.}}

\mysection{Data}{Data}
\change{
We show the 2D and 3D indicators used in \refFig{2D_3D_objects}.
}

\mysection{Speed}{Speed}
\change{We provide the speed of the bounding operation for all queries (point, ray, plane and box), dimensions and methods in \refTbl{SpeedAllQueries}. For each combination of dimension and query, we report the speed in milliseconds as an average over 10000 independent runs with 10 million randomly sampled queries. All methods are implemented as vectorized code to make full use of GPU acceleration on a workstation equipped with an RTX3090 GPU and Intel i9-12900K CPU.}

\change{Additionally, while the use of GPU-based libraries might incur a timing overhead, more traditional metrics like FLOPs are not suitable here either.
Take, \eg, the AABB inference test for point queries, x < x\textsubscript{max} \& x > x\textsubscript{min}, which is equal to zero FLOPs, as FLOPs don't count comparisons, logical operations or memory access, but still has a non-zero runtime.
The research of more robust timing metrics to compare non-neural to neural approaches is left to future work.}
\begin{table*}
    \centering
    \caption{\change{Query speed of methods in ms, averaged across 10,000 independent runs with 10 million samples per run.}}
    \label{tab:SpeedAllQueries}
    \begin{tabular}{lrrrrrrrrrrrr}
        \toprule
        &
        \multicolumn4c{2D}&
        \multicolumn4c{3D}&
        \multicolumn4c{4D}
        \\
        \cmidrule(lr){2-5}
        \cmidrule(lr){6-9}
        \cmidrule(lr){10-13}
        &
        \multicolumn1c{Point}&
        \multicolumn1c{Ray}&
        \multicolumn1c{Plane}&
        \multicolumn1c{Box}&
        \multicolumn1c{Point}&
        \multicolumn1c{Ray}&
        \multicolumn1c{Plane}&
        \multicolumn1c{Box}&
        \multicolumn1c{Point}&
        \multicolumn1c{Ray}&
        \multicolumn1c{Plane}&
        \multicolumn1c{Box}
        \\
        \midrule
        \method{AABox}& 1.03  & 2.08 & 2.08 & 2.08 & 1.27 & 2.58 & 2.58 & 2.58 & 2.04 & 4.21 & 4.21 & 4.20 \\ 
        \method{OBox} & 1.03  & 2.08 & 2.09 & 2.09 & 1.27 & 2.58 & 2.59 & 2.60 & 2.06 & 4.22 & 4.24 & 3.42 \\ 
        \method{Sphere} & 1.06  & 1.85 & 1.85 & 1.85 & 1.31 & 2.36 & 2.36 & 2.38 & 1.83 & 3.43 & 3.43 & 3.42 \\ 
        \method{AAElli} & 1.07  & 1.86 & 1.86 & 1.86 & 1.33 & 2.39 & 2.39 & 2.44 & 1.84 & 3.46 & 3.46 & 3.46 \\ 
        \method{OElli} & 1.07  & 1.87 & 1.86 & 1.87 & 1.33 & 2.39 & 2.45 & 2.43 & 1.84 & 3.46 & 3.47 & 3.47 \\ 
        \method{kDOP} & 12.42  & 14.36 & 14.37 & 14.35 & 10.61 & 41.27 & 41.40 & 41.39 & 14.06 & 54.84 & 54.87 & 54.66 \\ 
        \method{BVH} & 18.06  & 23.28 & 22.96 & 23.42 & 17.40 & 25.67 & 26.35 & 26.13 & 24.52 & 29.17 & 27.09 & 25.10 \\ 
        \method{OurkDOP} & 4.41  & 7.73 & 7.72 & 7.72 & 5.64 & 11.84 & 11.76 & 11.70 & 7.79 & 16.58 & 16.56 & 16.53 \\ 
        \method{OurNN} & 28.14 & 28.43 &  28.44 & 28.43 & 70.32 & 70.82 & 70.74 & 70.69 & 111.90 & 113.08 & 112.79 & 112.58 \\ 
        \method{OurNNEarly} & 24.55 & 29.26 & 32.39 & 30.71 & 35.97 & 46.30 & 74.70 & 50.90 & 64.86 & 96.15 & 135.08 & 101.95 \\ 
        \bottomrule
    \end{tabular}
\end{table*}

\mysection{Complexity}{Complexity}
From our results one can easily see that our method's advantage increases with dimensionality.
We visualize this in detail in \refFig{Complexity}.
This is because in concave shapes of natural high-dimensional signals, empty space grows much quicker than is intuitive.
High-dimensional space is mostly empty, 
\change{as the data tends to become sparse} \cite{aggarwal2001surprising}.
This is also why \acp{NN} excel as classifiers and best imagined in four dimensions, space plus time, where the space-time indicator of a moving objects is highly concave but mostly empty, and well represented by a \ac{NN} of low complexity: a bit of bending, and that is already much better than any box or sphere.

\myfigure{Complexity}{False positive rate in percentage (vertical axis) as a function of model complexity, \ie number of tuned parameters (horizontal axis in each subplot) for three different dimensions (subplots).
Different colors are different methods (\method{OBox}, \method{Sphere}, \method{OurNN}; baselines are constants with respect to complexity which remains fixed in each dimension).
Across dimensions, we see that \acp{FP} increase for baselines, while for ours, the rate (still function of model complexity itself) remains roughly constant (funnels).}

\bibliographystyle{ACM-Reference-Format}
\bibliography{paper}